\documentclass{aims} 
\usepackage{amsmath}
\usepackage{paralist}
\usepackage[misc]{ifsym}
\usepackage{epsfig} 
\usepackage{epstopdf} 
\usepackage[colorlinks=true]{hyperref}

\usepackage{graphicx,url,color}
\usepackage{lscape}
\usepackage{subfigure}
\usepackage{amsfonts}
\usepackage{mathrsfs}
\usepackage{mathrsfs}
\usepackage{amsmath}
\usepackage{amssymb}
\usepackage{float}
\usepackage{diagbox,multirow,makecell}
\usepackage{dsfont}
\usepackage{mathbbol}

\def\az{\alpha}  \def\bz{\beta}
    \def\dz{\delta}
    
\def\gz{\gamma}

\def\pz{\pi}

\def\rz{\rho}

\def\qd{\quad}
\def\qqd{\qquad}
\def\scr{\mathscr}
\def\le{\leqslant}
\def\ge{\geqslant}
\def\leq{\leqslant}

\def\nnd{\noindent}

\def\qed{ \hskip.5cm $\square$ \vskip.2cm}

\def\qed{\text{\quad $\square$}}

\def\de{\end{equation}}

\def\dear{\end{eqnarray}}
\def\lb{\label}

\def\den{\end{enumerate}}

\def\bbb{\mathbbold}

\allowdisplaybreaks[4]

\def\bbb{\mathbb}

\normalbaselineskip=\baselineskip \oddsidemargin=-0.25cm
\hypersetup{urlcolor=blue, citecolor=red}
\allowdisplaybreaks

\def\currentvolume{18}
 \def\currentissue{10}
  \def\currentyear{2025}
   \def\currentmonth{10}
    \def\ppages{2828-2852}
     \def\DOI{10.3934/dcdss.2025083 }

\newtheorem{theorem}{Theorem}[section]
\newtheorem{corollary}[theorem]{Corollary}

\newtheorem{lemma}[theorem]{Lemma}
\newtheorem{proposition}[theorem]{Proposition}

\theoremstyle{definition}
\newtheorem{definition}[theorem]{Definition}
\newtheorem{remark}[theorem]{Remark}

\newtheorem{example}[theorem]{Example}

\title[Hua-Chen New Theory of Economic Optimization]
{Hua-Chen New Theory of Economic Optimization} 

\author[BIN CHEN, YINGCHAO XIE, TING YANG and QIN ZHOU]{}

\subjclass{Primary: 91B99, 60J10, 15A99; Secondary: 15A72, 90B50.}
\keywords{Input-Output Model, Chen's Model, Chen's Invariant, Chen's Transform, Markov Chain.}

\thanks{All the authors are supported by National Natural Science Foundation of China 12471139, 12090011 and the Priority Academic Program Development of Jiangsu Higher Education Institutions.}

\thanks{$^*$Corresponding author: Qin Zhou}

\begin{document}
\maketitle

\centerline{\scshape
BIN CHEN$^{{\href{mailto:xzbchen@126.com}{\textrm{\Letter}}}1}$,
YINGCHAO XIE$^{{\href{mailto:6019820127@jsnu.edu.cn}{\textrm{\Letter}}}1}$,}
\centerline{\scshape
TING YANG$^{{\href{mailto:yangting@jsnu.edu.cn}{\textrm{\Letter}}}1}$
and QIN ZHOU$^{{\href{mailto:graceqinzhou@jsnu.edu.cn}{\textrm{\Letter}}}1}$}

\medskip

{\footnotesize
{ \centerline{$^1$Institute of Mathematics, School of
Mathematics and Statistics, }
\centerline{
Jiangsu Normal University, Xuzhou, China}}
} 

\medskip


\bigskip

 \centerline{(Communicated by Handling Editor)}


\begin{abstract}
Between 1957-1985, Chinese mathematician Loo-Keng Hua pioneered economic optimization theory through three key contributions: establishing economic stability's fundamental theorem, proving the uniqueness of equilibrium solutions in economic systems, and developing a consumption-integrated model 50 days before the immortal's passing of him. Since 1988, Mu-Fa Chen has been working on Hua's theory. He introduced stochastics, namely Markov chains, to economic optimization theory. He updated and developed Hua's model, and came up with a new model (Chen's model) which has become the starting point of a new economic optimization theory. Chen's theory can be applied to economic stability test, bankruptcy prediction, product ranking and classification, economic prediction and adjustment, economic structure optimization. Chen's theory can also provide efficient algorithms that are programmable and intelligent. {Stochastics} is the cornerstone of Chen's theory. There is no overlap between Chen's theory, and the existing mathematical economy theory and the economics developments that were awarded Nobel Prizes in Economics between 1969 and 2024. The distinguished features of Chen's theory from the existing theories are: quantitative, calculable, predictable, optimizable, programmable and can be intelligentable. This survey provides a theoretical overview of the newly published monograph \cite{5rw24}. Specifically, the invariant of the economic structure matrix, also known as the Chen's invariant, was first published in this survey.
\end{abstract}


\section{Introduction}

From the beginning of the last century to the 1960s, mathematics was in Hilbert's axiomatization era and the focus was on the construction of the foundations of various branches and on the solutions of important problems. In the 1960s, since most of the existing branches of mathematics have become mature, there was a need to open new fields, and so mathematics experienced a revolution and returned to the  Poincar\'{e} era, that is, returned to the era of integrating mathematics with various other disciplines such as  physics and biology. One representative of this revolution is the random field theory, developed by the Dobrushin's school at the intersection of probability and equilibrium statistical physics. Influenced by this, starting from 1980s, {Mu-Fa Chen}, along with his team {at} Beijing Normal University, built a new research direction, that is infinite-dimensional reaction-diffusion processes, as the {interaction} of probability theory and non-equilibrium statistical physics. {He published the research monograph \cite{c2004}.} In the late 1980's Chen and his team expanded their direction. On the theoretical side, they started a systematic study of stability
along with the associated convergence rate and spectrum theory. On the application side, they developed and revitalized Hua's economic optimization theory. Recently, there has been great progress in the applications of mathematics.  Mathematics has entered the front of many fields (such as weather forecasting, 3D printing, minimally invasive surgery, drones, robots and the booming artificial intelligence, etc) and triggered profound changes in many of these fields. All the applications mentioned above involve data collection, storage, rapid analysis, intelligent decision-making and rapid implementation. All these applications are supported by mathematical theory.

Applying mathematics to the economy, providing mathematical support for formulating national economic development plans, and establishing a set of in-depth methods to serve national economic development are the main parts of Hua's research direction in economic optimization, which is a model {of} interdisciplinary research. This is also the direction that {Chen} has been working in.

The materials on Chen's invariants for economic structure matrices are completely new and they have not been published elsewhere.

\section{Input-Output model and preliminaries}

\subsection{Input-Output model}

 In this survey, a ``product's vector'' (abbrev. d-vector) be regarded as an industry consisting of certain products, or a department consisting of some industries.
 We use the d-vector
 $$
 x=\left(x^{(1)},\cdots,x^{(d)}\right)
 $$
 to denote {the vector of the economic products we are concerned about, where $x^{(i)}$, $i=1, \cdots, d$,
 denotes the amount of the $i$-th product with fixed dimension.}

 To understand the current economic situation, we need the following data:
\begin{itemize}\setlength{\itemsep}{-0.8ex}
\item {{d-vector of last year}}: \;$x_0=\big(x_0^{(1)},\cdots,x_0^{(d)}\big)$; 
\item {{d-vector of this year}}: \;{$x_1=\big(x_1^{(1)},\cdots,x_1^{(d)}\big)$}; 
\item {{The structure matrix}}: \;$A_0=\big(a_{ij}^{(0)}: i,j=1,2,\cdots, d\big)$,
\end{itemize}
where the meaning of $A_0$ is as follows: to produce a unit of the $i$-th product, $a_{ij}^{(0)}$ units of the $j$-th products ($j=1, \dots, d$) are consumed. Thus we have
$$x_0^{(j)}=\sum_{i=1}^d x_1^{(i)} a_{ij}^{(0)},$$
in matrix form, the above can be written as
$$ x_0=x_1 A_0.$$
Conversely, given $x_0$ and $A_0$, to uniquely determine $x_1$, one needs $A_0$ to be invertible. Assume for now that all products produced are used for reproduction. This leads to the
\medskip
{\bf Idealized (no consumption) model}:
$$x_{n-1}=x_n A_{n-1},\; n\ge 1.$$
Hence
$$x_0=x_1 A_0=x_2 A_1 A_0=\cdots =x_n A_{n-1}\cdots A_0.$$
In the time-homogeneous case $A_n\equiv A$, we get the {\bf Input-Output model}:
\begin{equation}\label{2.1}
x_0=x_n A^n,\qqd n \ge 1.
\end{equation}
If we are only concerned with stability, that is, the limit behavior on the right side of the equation above as $n\to\infty$, we only need the asymptotic behavior of $A^n$, and $A$ do not need to be invertible. When $A$ is reversible, \eqref{2.1} is equivalent to
\begin{align}\label{2.2}
x_n=x_0 A^{-n}, \qqd n\ge 1.
\end{align}

\subsection{Two fundamental results of matrix theory}
To state the main results of Hua's theory, we need two fundamental results in matrix theory.
This section only involves two concepts: irreducibility and {aperiodicity}.
Let  $E=\{1, 2, \cdots, d\}$, $A=(a_{ij}: i, j\in E)$, $A^n=\big(a_{ij}^ {(n)}: i, j\in E\big)$.
In this survey, by a nonnegative matrix, we mean a matrix with all its entries {being} nonnegative. Positive matrix is defined similarly.
\medskip

\begin{definition}\label{def:irreducible}
A nonnegative matrix $A=(a_{ij})$ is said to be irreducible,
if for any pair $i, j\in E$, $i\neq j$, there exist distinct $i=i_1,\;i_2 ,\; \cdots,\; i_m=j$ such that
$$a_{i_1 i_2}>0,\; a_{i_2 i_3}>0,\; \cdots,\; a_{i_{m-1}i_m}>0.$$
\end{definition}

$A$ being irreducible  means that any pair of $i,j\in E$ are {\bf interconnected}, that is, there is a path from $i$ to $j$
and there is also a path from $j$ to $i$. If we put an edge between $i, j\in E$, $i\neq j$, whenever  $a_{ij}>0$, then we get a connect graph when $A$ is irreducible.
The following theorem is the most important result on nonnegative irreducible matrices, and it is the cornerstone of Hua's economic optimization theory. It is also a major difference between the Hua's theory and the current popular input-output theory.

\begin{theorem}[Perron-Frobenius 1907, 1912]\label{the:PF}
If $A$ is a nonnegative irreducible matrix, then its spectral radius $\rho(A)$\,$($in the aperiodic case$)$ is a positive simple eigenvalue of $A$, the corresponding {left- and right-eigenvectors are also simple and can be chosen to be positive}.
\end{theorem}

{Note that} $\rho(A)$ is the largest eigenvalue of $A$, and it is also called the {maximal} eigenvalue of $A$. Let $u$ (row vector) and $v$ (column vector) be the corresponding  {left- and right-} positive eigenvectors of $A$:
$$uA=\rho(A)u, \qqd Av=\rho(A)v.$$
$u$ and $v$ are called the {{maximal} left- and right-eigenvectors} of $A$. They are unique up to a multiplicative constant.

{Based on these facts, the triplet $(\rho(A), u, v)$ is called the three major characteristics of $A$.}
As seen above,  the row and column attributes of a vector can {be} often identified from the context and we will not explicitly indicate them. From now on, we assume that, for any $ i\in E$,
\begin{align}\label{2.3}
\big\{n\ge 1: a_{ii}^{(n)}>0\big\}\ne \emptyset.
\end{align}

\begin{definition}\label{def:aper}
The period of $i\in E$, $d_i$,  is defined to be the greatest common divisor of the elements in \eqref{2.3}.
If $d_i=1$, we say that {$i$} is {\bf aperiodic}.
When $A$ is irreducible, there is a common period for all $i\in E$, {which} is called the period of $A$ $($\cite{cm2021} Theorem $1.26$$)$. If the period is 1, then we say $A$ is aperiodic.
\end{definition}

If a nonnegative matrix $A$ has a positive diagonal entry, then it is {aperiodic}. The following results will be useful.

\begin{proposition}\label{pro:2.1}
If $A$ is a non-negative irreducible, {aperiodic} matrix, then exists a natural number $M\le (d-1)^2+1$ such that
$A^m$ is a positive matrix when $m\ge M$ $($\cite{m2000} {Example $8.3.4$ and Exercise $8.3.9$}$)$.
If all diagonal elements are positive,
the conclusion can be strengthened to $M\le d-1$ $($\cite{m2000}{ equation $(8.3.5)$}$)$.
\end{proposition}

Let $M_{\min}$ be the smallest $M$ satisfying the conclusion of the proposition above.

\begin{proposition}
If $A$ is a nonnegative irreducible and {aperiodic} matrix, then the modulus of every eigenvalue of $A$ different from $\rho(A)$ is less than $\rho(A)$.
\end{proposition}

Except Proposition \ref{pro:2.1}, the material in this section is classical and well known. For details, see \cite{cm2021} (Chapter 1), \cite{hua87}  or the entry ``Perron-Frobenius Theorem" in Wikipedia. The terminology used here is from the  theory of Markov chains. In matrix theory, ``irreducible" is called ``indivisible",  a ``irreducible, {aperiodic}, nonnegative square matrix"  is referred to as a ``{prime} square matrix".

We end this section with an example showing that the {invertibility} assumption of $A$ in the model is necessary.
Given a positive probability distribution $\pi$ (row vector) on ${\{0, 1\}}$.
Let  $A=\bbb{1}\pi$, where $\bbb{1}$ is a column vector with all components equal to $1$,
 then $A$ is irreducible and {aperiodic}, and the only stationary distribution is $\pi$: $\pi=\pi A$.
If $x_0=\pi$, then it is easy to check that any probability distribution $x_1$ satisfies the equation $x_0=x_1 A$.
{By assumption $A=\bbb{1}\pi$ and the condition that $x_1\bbb{1}=1$,
it follows that
       $$ x_1 A=x_1\bbb{1}\pi= \pi.$$
Hence, the equation $x_0=x_1 A$ has no solution once $x_0 \ne \pi$.}
This
indicates that this equation cannot completely determine an input-output system. The reason is that the rank of $A$ is $1$ and $A$ is not invertible.

\section{Hua's fundamental theorem of economic optimization}

From now on, unless explicitly stated otherwise, we will assume that $A$ is nonnegative, irreducible and {aperiodic}.

For a {d-vector $x$}, we write $\min_k x=\min\{x^{(1)}, \cdots, x^{(d)}\}$. For positive input {d-vectors $x$} ($x$ has no zero-component) and $y$, we define the quotient $y/x=(  y^{(1)}/x^{(1)}, \cdots,  y^{(d)}/x^{(d)})$. For a given input
d-vector $x$, {let $y$ be the d-vector $xA$, then} the corresponding quotient is $y/x=xA/x$.
The following lemma gives a variational formula for the {maximal} eigenvalue $\rho(A)$ of $A$ using $xA/x$.

\begin{lemma}[Collotz-Wielandt (C--W) formula]\label{le:CW}
 If $A$ is a nonnegative irreducible matrix, then
\begin{align}\label{3.1}
\max_{x>0}\min_k \frac{xA}{x}=\rho(A)=\min_{x>0}\max_k \frac{xA}{x},
\end{align}
{where}
{$\max_k \frac{xA}{x} =\max_k \frac{(xA)^{(k)}}{x^{(k)}}. $}

  The equalities above remain valid if $xA$ is replaced by $Ax$.
Two {equalities} hold simultaneously if and only if $x$ is the corresponding {maximal} eigenvector.
\end{lemma}

In applications, we often use the C-W formula \eqref{3.1} instead of the Rayleigh entropy method to estimate $\rho(A)$. This formula is universal, that is, for each positive vector  $x$, there are lower and upper {bound} estimates:
$$ \min_k \frac{xA}{x}\leq \rho(A )\leq \max_k \frac{xA}{x}.$$

If $x_0$ and $x_1$ are the input and output indices respectively, we define $\min_kx_1/x_0$ to be the development rate of {economy}. The following result is Hua's fundamental theorem in  economic optimization.

\begin{theorem}[Hua's Fundamental  Theorem in Economic Optimization]\label{the:hua}
\quad Let\linebreak $x_{n-1}=x_{n}A$,\;$n\!\ge\!1$, $\rho(A)$, $u$, $v$ be the {maximal left- and right-eigenvectors} of ~$A$. Then the following conclusions hold.
\begin{itemize} \setlength{\itemsep}{-0.8ex}
\item[{\rm (i)}] The optimal input $x_0=u$, which gives the optimal development rate $1/\rho(A)$.
\item[{\rm (ii)}]
If $A$ is aperiodic and~$x_0\ne u$, then the economic system will inevitably  collapse, that is, the collapse time
$$ T_{x_0}^+=\inf\{n\!: x_n\;\text{\rm contains some negative component}\}<\infty. $$
\end{itemize}
\end{theorem}

\begin{remark}\setlength{\itemsep}{-0.4ex}
This theorem says that $u$ is optimal from two different perspective.
\begin{itemize}
\item
From the perspective of economic development rate,
the optimal input is the {left-eigenvector} ~$u$, that is,~$x_0=u$;
\item
In order for the economic system to never collapse, the optimal input must also be the {left-}eigenvector~$u$, That is,~$T_{x_0}^+={\infty}\Longrightarrow x_0=u$.
\end{itemize}
\end{remark}

\vspace{3mm}
\nnd {\bf Proof:} (i) The slowest development rate of the economy is
$$
\min_k \frac{x_0A}{x_0}.
$$
To attain the optimal development rate, we need to look for $x_0$ so that the quantity above achieve maximal. It follows from
Lemma \ref{le:CW} that the maximal is achieved when $x_0$ is the {maximal} {left-}eigenvector $u$ of $A$ and the optimal development rate is $1/\rho(A)$.

(ii) The following is the proof given by  Chen in September of 1989 in the case when $A$ is a transition probability matrix.
Assume $A=P$ is a transition probability matrix. Then $P{\bbb{1}}={\bbb{1}}$,
that is, the {maximal} right-eigenpair of $P$ is $(1, {\bbb{1}})$,
and the optimal development rate is $1$.

Assume $P$ is irreducible and aperiodic. Then the ergodic theorem of Markov chains
(\cite{cm2021} Theorem 1.19)
$$P^n\to {\bbb{1}} \pi, \quad n\to{\infty},$$
where $\pi=(\pi^{(1)},\cdots,\pi^{(d)})$ ($\pi^{(i)}>0,i=1,\cdots, d$) is the invariant probability measure of~$P$.

Assume $\mu_0=\mu_nP^n, \forall n\ge 1$, $\mu_0\ge 0$ satisfies $\mu_0\bbb{1}=1$. It follows from $P{\bbb{1}}={\bbb{1}}$ that
$$
\mu_0{\bbb{1}}=\mu_nP^n{\bbb{1}}=\mu_n{\bbb{1}}=1,~~~~\forall n.
$$
Combining this with the nonnegative boundedness of $\{\mu_n\}$,  we get that  there exists a convergent subsequence~$\{\mu_{n_k}\}_{k\ge1}$. If $\mu_{n_k}\to {\bar\mu}$, then ${\bar\mu}{\bbb{1}}=1$.
Therefore
$$\mu_0=\lim_{k\to{{\infty}}}\mu_{n_k}P^{n_k}={\bar \mu}{\bbb{1}}\pi=\pz.\qed$$

For a general structure matrix $A$, is it possible to transform $A$ into a transition probability matrix $P$ and use it to prove
Hua's fundamental theorem? This leads to {\bf Chen's new theory of economic optimization}.

It is well known that {equilibrium} is the core element of the economy.
If the balance is destroyed, it may not be restored for a hundred years.
In the 1990s, virtual economy developed rapidly due to the rise of the internet  and countless financial products emerged.
When most of these virtual products defaulted,  the financial world collapsed quickly, causing a series of collapses in the real economy.  When the financial crisis broke out in 2008, Iceland declared national bankruptcy.

People often say that ``balance is temporary, and imbalance is the norm". In a rigorous mathematical framework,
balance or equilibrium  means a certain ratio between products.
The  equilibrium solution is unique up to a multiplicative constant.
In practice, it is important to allow this free factor.
No matter how many products an economic model has, the equilibrium solution is just a ratio between products.
The importance of the equilibrium solution lies in that it is the only stable solution of the economic system, that is, starting from it,
we get $x_1=x_0/\rho(A)$, the d-vector of each step is $1/\rho(A)$ times that of the previous step.  $1/\rho(A)$ is uniquely determined by the economic system and it is the  optimal development rate.
 In other words, the output of each step is an equilibrium solution, so it is also called a stable solution.
 This is the first major result of Hua's theory. Most of the results on economic optimization before Hua were existential rather than constructive, Hua's results are computable.


Hua's most surprising result is part (ii) of Theorem \ref{the:hua}. In the ideal situation without consumption, if the system does not start from the equilibrium solution, then the economy will inevitably have products with different signs at a certain step (year), that is, the economy will collapse or go bankrupt. We will prove that the result remains true in the general model with consumption.

The following is the simplest model derived from Wassily Leontief that Hua has simplified.

\subsection*{The ancient economy of men farming and women weaving}

In this economy model, there are only two sectors (products): agriculture and manufacturing.
We fix product units;
such as buckets and liters in agriculture, feet and yards in manufacturing, etc.
The following table gives the rules of economic operation in this model.

\begin{center}{\begin{tabular}{l|c|c}
\Xhline{1pt}
\diagbox{Output}{\;Consumption\;}{\;Input\;} & Agriculture & Manufacturing \\
\hline
Agriculture & 0.25 & 0.14 \\
\hline
Manufacturing & 0.4 & 0.12 \\
\Xhline{1pt}
\end{tabular}}\end{center}\medskip

\nnd The first row in the table says that to produce~1 unit of agricultural product, $0.25$ unit of agricultural product
and $0.14$ unit of manufacturing product are consumed.
Similarly, the second row say that to produce 1 unit of manufacturing product, $0.4$ unit of agricultural product and $0.12$ unit of manufacturing product are consumed.

The matrix composed of the four numbers in the table is called
the ``consumption coefficient matrix" or the ``structure matrix".

\vspace{-0.25cm}
\begin{align*}
A=\frac{1}{100}\begin{bmatrix} 25 & 14 \\
40 & 12\end{bmatrix}
=\begin{bmatrix} 0.25 & 0.14 \\
0.4 & 0.12\end{bmatrix},
\end{align*}
%
Write the output d-vector of the $n$-th year as $(x_n, y_n)$, where $x_n$ is the output of agriculture in the $n$-th year, and $y_n$ is the output of manufacturing in the $n$-th year. Write the initial input index as $(x_0, y_0)$. How to choose the input appropriately is the core of  this theory.

The first contribution of Hua in economic optimization is finding out that the input-output table has a unique (up to a multiplicative constant) equilibrium solution or stable solution, which is the {maximal}  {left-}eigenvector of $A$.
 \begin{align*}
 u &\!=\!\text{\rm (agriculture,\; manufacturing)} \\
 &\!={\!\bigg(\frac 5 7 \big(\sqrt{2409}+13\big), \;20\bigg)
 \dot= \!~~(44.34397483,\; 20),}
 \end{align*}
where $\dot=$ represents approximation.
Here uniqueness is up to a multiplicative constant. In practice, any multiple of $u$ can be used.
This is of course necessary in practice:
The number of products used by a small enterprise and the number the same order of magnitude.
In the economic model we are studying,  the input is allowed to differ by a multiplicative constant.
 {Hua} proved that if $x_0=u$ is selected as the input, {then} the economic system will always remain in balance and growth rate will be optimal.
He called this method the ``positive eigenvector method" (the  ``positive eigenvector" here  is the equilibrium solution $u$ mentioned above), which is a ``proportional, high-speed development" method.
 Hua's most  brilliant conclusion is that if $x_0\ne u$ is selected as the input, then in a certain year, the d-vector of the economy will definitely have components with different signs. The phenomenon is commonly referred to collapse
or bankruptcy.  We will use $T$ to denote the first year that this situation occurs and call $T$ the collapse time or bankruptcy time. We will call the few years before $T$ crisis times. If the collapse time is very far away, we do not need to worry too much. The surprising thing is that the system is very sensitive.  Starting from the initial input $(x_0, y_0)$, we can get the
output $\{(x_n, y_n)\}_{n\ge 1}$ successively. We give two examples here and find out their bankruptcy  and crisis times.
\medskip

\begin{example}\label{exam:3.1}
Set the initial input value
to be the equilibrium solution rounded to $3$ decimal places:~$(x_0, y_0)=(44.344, 20)$,
and the result is shown in Figure~\ref{fig:exam3.1} below. The blue line represents agriculture, and the red line represents manufacturing. As can be seen from the figure, the first $7$ years $($steps$)$ ran well, but the increase in the $6$th and $7$th steps was a bit too fast, leading to a collapse in the next step,  $T\!=\!8$. For this initial value, the five-year plan is feasible.
\end{example}

\begin{figure}[h]
\begin{center}
\includegraphics[width=6.0cm,height=6.0cm]{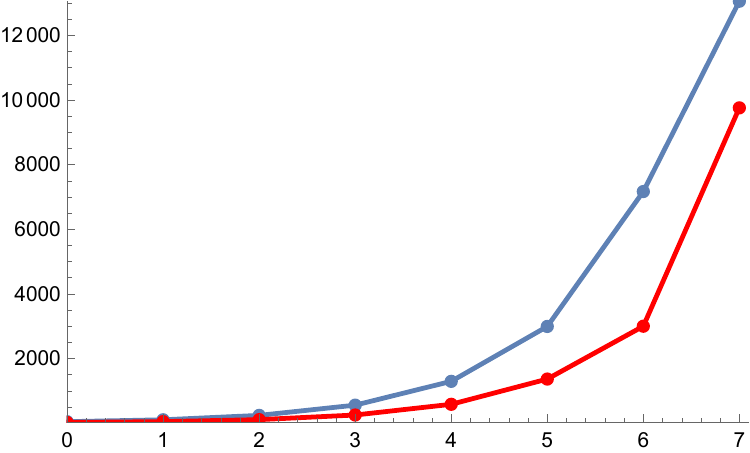}
\includegraphics[width=6.0cm,height=6.0cm]{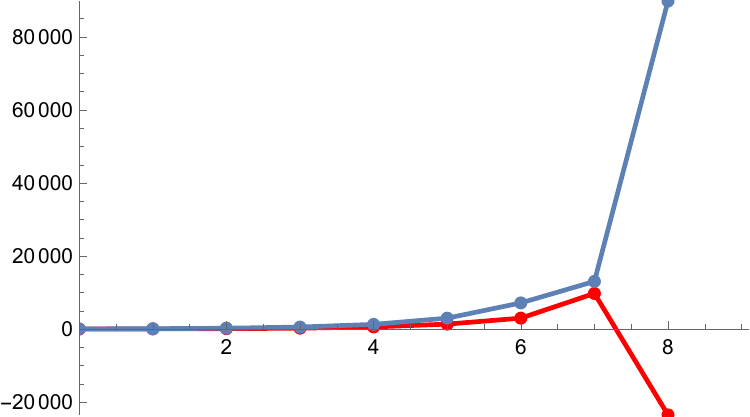}
\end{center}
\caption{}
\label{fig:exam3.1}
\end{figure}

\begin{figure}[h]
\begin{center}
\includegraphics[width=6.0cm,height=6.0cm]{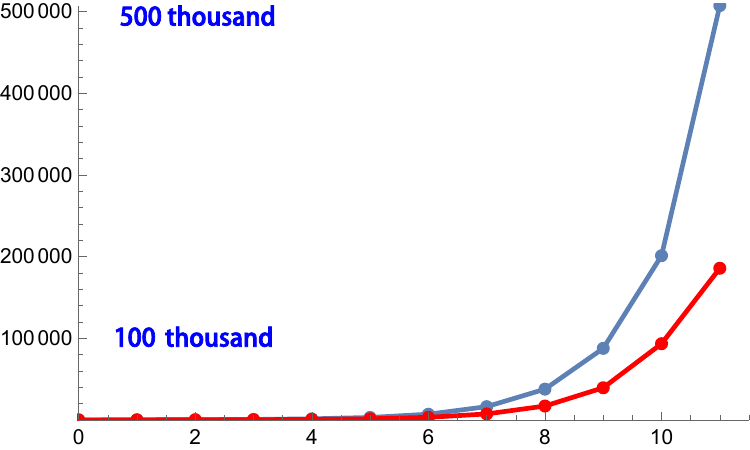}
\includegraphics[width=6.0cm,height=6.0cm]{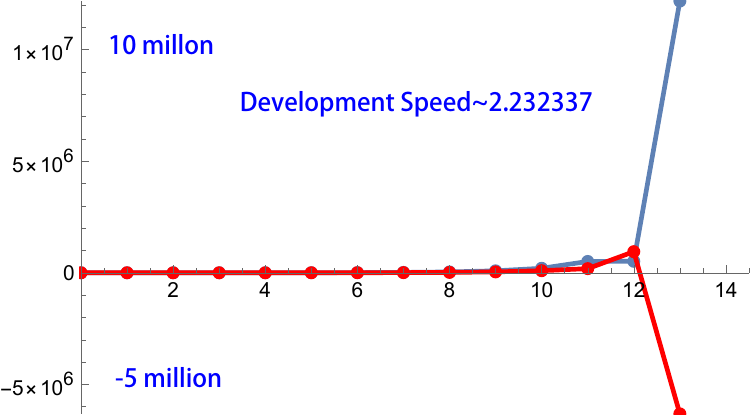}
\end{center}
\caption{}
\label{fig:exam3.2}
\end{figure}

\begin{example}\label{exam:3.2}
Set the initial input value to be the equilibrium solution rounded to $8$ decimal places:~$(x_0, y_0)=(44.34397483,\; 20)$, as shown in Figure \ref{fig:exam3.2},
the collapse time is $T=13$. For this example, the precision used is already very high,
which is not easy to achieve in practice, but it can only run for  $12$ years at most.
In the $13$th year, the maximal value of the upper curve reaches ten of millions,
and the minimum value of the lower curve reaches negative millions.
It is easy to see that the crisis time is  $10$ or  $11$.
\end{example}

\nnd
The left graphs in Figure \ref{fig:exam3.1} and Figure \ref{fig:exam3.2} above stop at the crisis time, and the operation before this is relatively stable.
 Go forward one or two steps, the system collapses,  forming the right graph above.
The scale of the right graph is compressed a lot, and this makes the right graph very different the left one.
The two graphs do not have the same order of magnitude.
These two examples show the sensitivity of the system is beyond imagination.

It can be seen from the two graphs above that both products grow very fast intially.
Most of the right figure is covered by the last big jump,
and the growth of the previous steps is almost invisible.
This method is not feasible for a larger system since hundreds of iterations might be needed and our present day computer
are not powerful enough to handle the computations. We need to find a new way.

\section{{Chen's Model}}

In 1957,  Hua gave some lectures in the Institute of Mathematics of the Chinese Academy of Sciences on the input-output method, popular in the economic community at that time,  established by Wassily Leontief in 1936 (Nobel Prize winner in Economics in 1973).
Leontief removed the traditional basic requirement of distinguishing between production data and consumption data,  simplified the conventional model and successfully applied the new model in practice.
From the Perron-Frobenius Theorem \ref{the:PF}, {Hua} realized that production data and consumption data should be clearly differentiated.
In 1958--1959, in the textbook ``Introduction to Advanced Mathematics" (Volume 4) written for students of the Mathematics Department at the University of Science and Technology of China, he gave a new proof of the aforementioned theorem.
The core part of Hua's model is keeping the tradition of clearly separating production data and consumption data.
For models involving consumption, new methods are needed. This was the a major challenge in our input-output model actually.
Between 1984 and 1985,   Hua tried several methods and published more than 10 research {reports} in the {``Chinese Science Bulletin"}.  The key point was that each year, a portion of the increase in production should be allocated for consumption.
 Hua proposed  two approaches for this. In papers \cite{hua84-3} and \cite{h1984}, he used
\begin{align}\label{4.1}
x_n-\xi_n=x_{n+1}A,
\end{align}
where $\xi_n$ is the consumption; in the paper \cite{hua85-2},
he denoted the left-hand of the equation above by $y_n$,  and hence \eqref{4.1} becomes
$y_n=x_{n+1}A$. Suppose that, for some  $\alpha\in (0, 1)$,  an $\alpha$ portion of the increment in production in the $n$-th
year, i.e., $\alpha(x_n-y_{n-1})$, is used for consumption, then the amount that d-vector that can be used for reproduction in the $(n+1)$-th year is
\begin{align*}
y_n&=x_n-\alpha(x_n-y_{n-1})\\
&=(1-\alpha)x_n+\alpha y_{n-1}\\
&=(1-\alpha)y_{n-1}A^{-1}+\alpha y_{n-1}\\
&=y_{n-1}[(1-\alpha)A^{-1}+\alpha I].
\end{align*}
Thus we obtain the model
\begin{align}\label{4.2}
y_n=y_{n-1}B, ~~~~ n\ge 1,
\end{align}
where the matirx $ B=(1-\az)A^{-1}+\az I$ is no longer
nonnegative and $\az\in(0, 1)$ the consumption ratio. In this case the economic growth rate is $(1-\az)\rho(A)^{-1}+\az$.

As far as we know, Model \eqref{4.2} was in use until around October of 2021.
However,   Hua was not satisfied with any of the aforementioned models.
About 50 days before the immortal's passing of him, he removed all previous research {reports} from the references of his manuscript \cite{hua87} and proposed a new model, that is,
in \eqref{4.1}, set
$$
\xi_n=\gz (x_{n+1}-x_n),
$$
where $\gz\in(0, 1)$ is the consumption ratio, and so
$$
x_n=x_{n+1}A+\gamma(x_{n+1}-x_n)=x_{n+1}(A+\gamma I)-\gamma x_n.
$$
Hence we have
$$x_n=x_{n+1}A_{\gz},\qd A_{\gz}=\frac{A+\gz I}{1+\gz},\qd \rho(A_{\gz})=\frac{\rho(A)+\gz}{1+\gz}.$$
This returns to the model  with no consumption and with nonnegative structure matrix $A_{\gz}$:
\begin{align}\label{4.3}
x_0=x_n A_{\gz}^n.
\end{align}
Unfortunately, the Model \eqref{4.3},  Hua's ``Last Thought" remained dormant for 37 years before it was  awakened by  Chen \cite{c2022}.

The development rate of the  economic system corresponding to Model \eqref{4.3} is $1/\rho(A_\gamma)$ when $\rho(A)\ge 1$, the economic system is abnormal and further research is required. Thus we assume that $\rho(A)<1$. Then the corresponding growth rate is
$$
\frac{1}{\rho(A_{\gz})}-1=\frac{1-\rho(A)}{\gz+\rho(A)}~\big\downarrow~\frac{1-\rho(A)}{1+\rho(A)}>0,~~~\gz\uparrow 1.
$$
This shows that,  for any $\gamma\in(0,1)$,  Model \eqref{4.3} always has a positive growth rate, which is inconsistent with reality.
Therefore, Model \eqref{4.3} has a flaw. Noting that
$$
\frac{1-\rho(A)}{\gz+\rho(A)}~\big\downarrow~0~\Longleftrightarrow~\gz\uparrow\infty,
$$
%
 Chen proposed the following model. Take
$$\gz=\frac{\az}{1-\az},$$
where $\az\in(0,1)$ is the consumption parameter,
and so $\gamma\in(0,\infty)$. This completely different from the assumption $\gamma\in(0,1)$ in Model \eqref{4.3}. In  practice, $\gamma>1$ may occur. At last, the new model becomes
\begin{align}\label{4.4}
x_0=x_n A_{\az}^n,\qd\qd n\ge 1,
\end{align}
where
$$A_{\az}=(1-\az) A+\az I,~~~\rho(A_{\az})=(1-\az)\rho(A)+\az.$$
Obviously $A_{\az}~(\az\!<1)$  has the same {maximal} {left- and right-eigenvectors} $u$
and $v$ as $A$. Model \eqref{4.4} is called {\bf Chen's model}.

The matrix $A_{\az}$ in Chen's model can be constructed in two steps:
$$
A\overset{(\text{i})}\Longrightarrow (1-\az)A\overset{(\text{ii})}\Longrightarrow (1-\az)A+\az I=A_{\az}.
$$
\begin{itemize}\setlength{\itemsep}{-0.8ex}
\item[(i)]
Since a portion of the the resources is used for consumption, $\az$ times of the structure matrix $A$ is deducted from $A$.
Thus, only  $1-\az$ times of $A$ is actually invested in reproduction.

\vspace{0mm}
\item[(ii)]
The deducted part is not used for reproduction, so the off-diagonal elements of the deducted part of the new structure matrix
are all 0; each product still contributes an $\az$ portion to the new structural matrix, to be used for consumption.
\end{itemize}

\section{Key transform of non-negative matrices -- Chen's transform}
  As mentioned before,  when the structure matrix $A$ is a transition probability matrix,  Chen has given a proof of Hua's fundamental theorem. How do we transform a general structure matrix into a transition probability matrix?
Chen has been working on this topic for a long time.

The first step is figure out what needs to be done to transform $A$ into a transition probability matrix $P$.
 We know that, for transition probability matrix $P$, its {maximal} {left-}eigenvector $\pi$ is the stationary distribution of $P$.
 Markov chain theory tells us that, under the irreducibility condition, the stationary distribution of $P$ is unique.
%
%
%
%
%
%
%

\begin{definition}
{The nonnegative matrix having sum 1 of each row is called a transition
probability matrix.}
\end{definition}

\begin{lemma}\label{le:max1}
{The maximal eigenvalue of a transition probability matrix equals one.}
\end{lemma}

\nnd{\bf Proof:}
{Let~
$\|x\|_{\infty}=\sup_k |x^{(k)}|$, $\|P\|_{\infty}=\sup_{x\ne 0} |Px|_{\infty} /{\|x\|_{\infty}}$.
Obviously 1 is an eigenvalue of $P$, and its corresponding eigenvector is ${\bbb{1}}$, we can get that its maximal eigenvalue has a lower bound of $1$.
On the other hand, according to the definition of the eigen-equation:
$$Px=\lambda x.$$}
{Because
$$(Px)(k)=\sum_{j} p_{kj}x^{(j)}\le \|x\|_{\infty}\sum_{j} p_{kj}=\|x\|_{\infty},$$
hence
$\|Px\|_{\infty}|\le \|x\|_{\infty}$.
We get the maximal eigenvalue of $P$
$$\lambda_{\max}=\sup_{x\ne0}\frac{\|Px\|_{\infty}}{\|x\|_{\infty}}
\le 1. \qed $$}

\begin{remark}
{If $P$ is in addition irreducible, then the conclusion of Lemma $\ref{le:max1}$ can be derived directly from the C-W formula.}
{In fact, since $1$ is the eigenvalue of ~$P$, and $\bbb {1}$ is the corresponding eigenvector, applying the C-W formula to the vector~$\bbb {1}$ we can get $\lambda_{\max}\le 1$.}
\end{remark}

For a positive vector $w$, we use $D_w$ to denote the diagonal matrix with diagonal entries given by the components of $w$. For any vector $x$, we define the component product $w\odot x=(w^{(1)}x^{(1)},\cdots,w^{(d)}x^{(d)})$ and
\begin{align}\label{5.1}
A_w\hat=D_w^{-1}\frac{A}{\rho(A)}D_w.
\end{align}
Then we have the following result.

\begin{theorem}[Chen 1989, 1992, 2022]\label{the:chen1989}
{From} \eqref{5.1}, the following conclusions hold:

$\mathrm{(i)}$
$A_w$ is a transition probability matrix $P$ if and only if $w=v$;

$\mathrm{(ii)}$
The {maximal} {left-eigenvector} of $P$ is $\mu:=u\odot v$;

$\mathrm{(iii)}$
$\pi:=u\odot v/(uv)$ is a stationary distribution of $P$: $\pi=\pi P=\pi P^n$.
\end{theorem}

\nnd{\bf Proof:}
 Without loss of generality, we assume $\rho(A)=1$.

{ (i)} {From} \eqref{5.1}, we know that
\begin{eqnarray*}
A_w{\bbb{1}}=D^{-1}_wAD_w{{\bbb{1}}}=D^{-1}_wAw \stackrel{?}{=}  \bbb{1}.
\end{eqnarray*}
Conclusion (i) holds if and only if $Aw=D_w{{\bbb{1}}}$, i.e. $Aw=w$. Therefore $A_w$ is a transition probability matrix if and only if $w=v$.

(ii) It follows from (i) that
\begin{align}\label{5.2}
P=D_v^{-1}\frac{A}{\rho(A)}D_v.
\end{align}
%
Thus we have
$$
\mu P=u\odot v P=u\odot v D^{-1}_vAD_v=uAD_v=uD_v=u\odot v=\mu,
$$
That is, $u\odot v$ is {the {maximal} left-eigenvector of $P$}.

$\mathrm{(iii)}$ Define $\pi=u\odot v/(uv)$.
It is easy to verify that $\pi$ is the stationary distribution of $P$.
Combining
$$
P{\bbb{1}}=D^{-1}_vAD_v{\bbb{1}}=D^{-1}_vAv=D^{-1}_vv={\bbb{1}}
$$
with Lemma \ref{le:max1}, we get that  $\rho(P)=1$ and the {maximal right-eigenvector} of $P$ is ${\bbb{1}}$.\qed

\begin{remark}{}
\begin{itemize}
\item[$\mathrm{(1)}$]
The sufficiency part of Theorem $\ref{the:chen1989}${$\mathrm{(i)}$}  was obtained by  Chen in $1989$ and published in $1992$.  The necessity part
was proved in $2022$. Theorem $\ref{the:chen1989}${$\mathrm{(ii)}$} and $\mathrm{(iii)}$ were also proved by Chen in $2022$.

\item[$\mathrm{(2)}$]
We will call \eqref{5.2} {\bf{Chen's transform}} of $A$, which was given by  Chen in $2022$.
\end{itemize}
\end{remark}

%
%
%
%
%
%

{The equilibrium $\mu=u\odot v$ plays an important role in the present theory, as shown in the next section.}

Now we present Chen's proof of Hua's fundamental theorem in the case of general structure matrices.

\nnd{\bf Proof of Hua's fundamental theorem:}
It follows from Theorem \ref{the:chen1989} (i) that
\begin{align}\label{5.3}
D^{-1}_v\left(\frac A{\rho(A)}\right)^nD_v=P^n.
\end{align}
Thus
$$
\left(\frac A{\rho(A)}\right)^n=D_vP^nD^{-1}_v.
$$
Assume that $\{x_n\}_{n\ge 0}$ satisfies $(x_0D_v){\bbb{1}}=1, x_0=x_nA^n$ and $x_n\ge0 ~(n\ge0$).
By the definitions of $u$ and $v$ and the assumption $(x_0D_v){\bbb{1}}=1$, we know that, up to a multiplicative constant, $x_0=u$.
Note that
\begin{align}\label{5.4}
x_0D_v\!=\!x_nA^nD_v\!=\!\left[\rho(A)^nx_nD_v\right]D^{-1}_v\!\left(\frac A{\rho(A)}\right)^n\!\!D_v\!=\!\left[\rho(A)^nx_nD_v\right]P^n.
\end{align}
Denote $y_n=\rho(A)^nx_nD_v$,
\eqref{5.4} can be rewritten as
$$
y_0=y_nP^n,n\ge 0
$$
and $y_0{\bbb{1}}=1$ ($n\ge 0$).
Combining Theorem \ref{the:chen1989} and the proof of Hua's fundamental theorem in the special case of $A=P$, we can get $y_0=u\odot v/(uv)$,
that is, $x_0=u/(uv)$.
In order to make all $x_n$ nonnegative, the initial value {must} $u$.
\qed

For given $P$ and $\mu_0$, define $\mu_0=\mu_nP^n$, $n\ge 1$. $\{\mu_n\}_{n\ge0}$ is called the iterative sequence of $\mu_0$ under $P$. Similarly, {the} iterative sequence $\{x_n\}_{n\ge 0}$  of $x_0$ under $A$ is defined by $x_0=x_nA^n$, $n\ge 1$. The following result says that the equivalence of these two iterative sequences.

\begin{theorem}\label{the:mux}
$\{\mu_n\}_{n\ge0}$, $\{x_n\}_{n\ge 0}$
and $v$ satisfy the following identities:
\begin{align}\label{5.5}
\mu_n=\rho(A)^nx_n\odot v,\,n\ge 0,
\end{align}
\vspace{-0.5cm}
\begin{align}\label{5.6}
x_n=\rho(A)^{-n}\mu_n\odot v^{-1},\,n\ge 0.
\end{align}
\end{theorem}

\nnd{ \bf Proof:}
 Let $\{x_n\}_{n\ge 0}$ be the iterative sequence of $x_0$ under $A$. Then \eqref{5.4} holds.
It follows from $x_0D_v=\rho(A)^0x_0\odot v$ that $\mu_n:=\rho(A)^nx_n\odot v (n\ge 0)$, { satisfies $\mu_0=\mu_n P^n, n\ge 1$}. That is, \eqref{5.5} holds.
\eqref{5.6} can be proved similarly.
\qed

Multiplying both sides of \eqref{5.5} by the vector ${\bbb{1}}$ on the right {side} yields
$\rho(A)^nx_n v=\mu_n{\bbb{1}}$.
In particular, if we take $x_0=u$, then $\mu_0=u\odot v$,
the normalization condition is
$uv=\mu_0{\bbb{1}}=1=\pi{\bbb{1}}.$
Note that $\mu_n$ and $x_n$ are related like in \eqref{5.5}.

Next we consider Chen's transform $P_{\az}$ of the structure matrix $A_{\az}=(1-\az)A+\az I$ in the case with consumption.
Obviously, the maximal eigenvalue of $A_{\az}$ is $\rho(A_{\az})=(1-\az)\rho(A)+\az$.
Note that $A_{\az}$ and $A$ have the same {maximal right-eigenvector} $v$, thus Chen's transform of $A_{\az}$ is
\begin{align}\label{5.7}
P_\az=&{D_{v}^{-1}}\frac{A_{\az}}{\rho(A_{\az})}D_v\nonumber\\ =&{\rho(A_{\az})}^{-1}\big[(1-\az)D_{v^{-1}}AD_v+\az I\big]\nonumber\\ =&{\rho(A_{\az})}^{-1}\Big[(1-\az)\rho(A){D_{v}^{-1}}\frac{A}{\rho(A)}D_v+\az I\Big]\nonumber\\ =&(1-\bz_{\az})P+\bz_{\az} I,
\end{align}
where $\bz_{\az}=\az/ \rho(A_{\az})$.

\eqref{5.7} shows that $P_{\az}\,(\az\!<\!1)$ has the same
{maximal left- and
right-eigenvectors} $u\odot v$ and ${\bbb{1}}$.
One can similarly prove the equivalence of the  two iterative algorithms for $A_\alpha$ and $P_\alpha$.

{To conclude this section, let us illustrate the power of Chen's transform. For this, we return to the
stability testing of Examples \ref{exam:3.1} and \ref{exam:3.2}, using $P$ instead of $A$. The results are given by Figures \ref{fig:5.1}
and \ref{fig:5.2} below. Certainly, the initials for $P$ should be different {from} that of $A$.}


\begin{figure}[h]
\begin{center}
  \hspace{-1.5cm}
  \includegraphics[width=12.5cm,height=6.5cm]{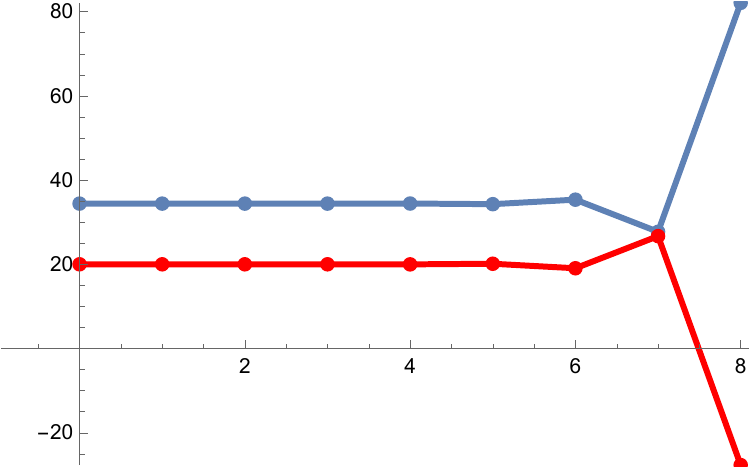}
  \caption{ New figure corresponding to Figure \ref{fig:exam3.1}. Initial ~$(34.41181135,\;20)$,\;$T=8$}\label{fig:5.1}
  \end{center}
\end{figure}

\begin{figure}[h]
\begin{center}
  \includegraphics[width=12.5cm,height=6.5cm]{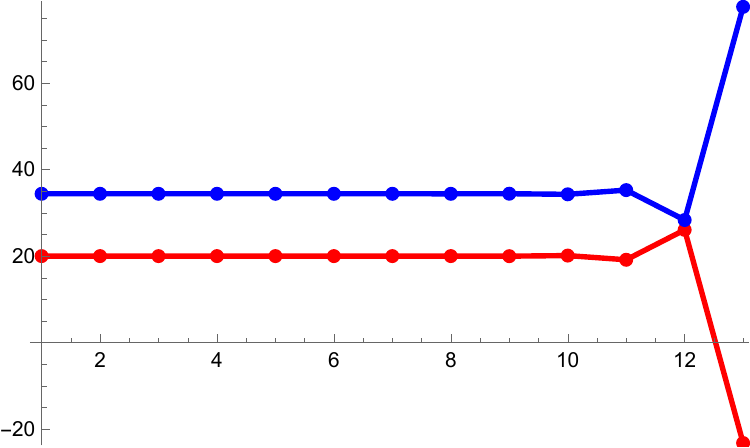}
  \caption{ New figure corresponding to Figure \ref{fig:exam3.2}. Initial ~$(34.41179182,\;20)$,\;$T=13$}\label{fig:5.2}
  \end{center}
\end{figure}

{Very unexpected, the stability for $P$ and $A$ are completely the same (i.e., having the same collapse time and the same
collapse place (at the same product)). Surprisingly, the picture generated by~$P$ is much better than the one generated by~$A$. Therefore, we use $P$ only in practice for stability testing.}

\section{{Product ranking and classification}}

The second part of Theorem \ref{the:hua} is Hua's main contribution to economic optimization, and as far as we know, it has never appeared before.
It says that, if $x_0\ne u$, {then} {the} system will collapse at an exponential rate.
 Therefore, it is important to understand the classification of products in the economic system: pillar products, intermediate products and weak products,
 because the system often collapses at some weak products.

For a nonnegative and irreducible matrix $A$,
the {maximal} {left-}eigenvector $u$ {contains only} two characteristics of $A$.
But the {maximal} {left-}eigenvector $\mu=u\odot v$ of $P$ obtained from $A$ via Chen's transform contains all three characteristics of $A$.
As mentioned before, $u\odot v$ represents the true total value of the products and has a unified dimension. Therefore, compared with using the {maximal} {left-}eigenvector $u$ of $A$,  it is more scientific to use the
 {maximal} {left-}eigenvector $\mu$ of $P$ to ranking products.
{Furthermore, as we have seen from the last section that using $P$ or $A$, we have completely} the same stability, but the amplitude of oscillation of {the} former is less than {the one of } the latter.

\begin{figure}[h]
\begin{center}
\includegraphics[width=13cm,height=8.6cm]{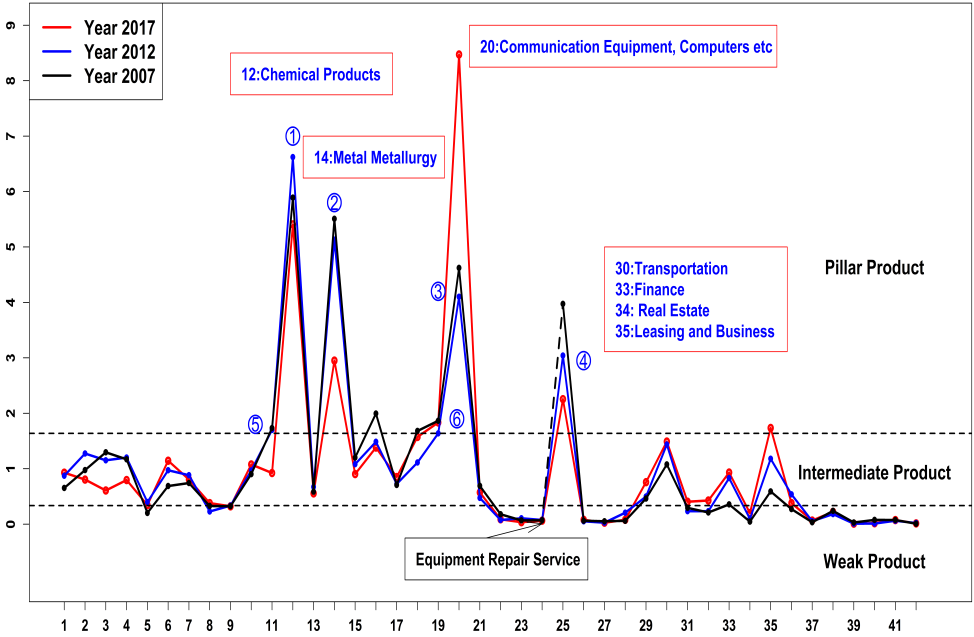}
\end{center}
\caption{The equilibriums diagram of Year 2017's 42 products, Year 2012's 42 products and Year 2007's 41 products. }
\label{fig:exam6.1}
\end{figure}

To illustrate the significance of product ranking, we give some practical examples.
Figure \ref{fig:exam6.1} gives the rankings of 42 products in China's input-output tables in 2017 (red), 2012 (blue) and 2007 (black).
The National Bureau of Statistics in China compiles an input-output table every 5 years, so Figure \ref{fig:exam6.1} contains detailed information about China's economic situation in a 15 year period. For details, see  Chapter 4 of \cite{5rw24}.
For ease of observation, the vertical axis represents  multiples of the equilibrium solution of the transition probability matrix $P$.
The three curves are surprisingly similar.
Before we analyze the figure, we should inform our readers that, in the 2007 input-output table, we did a numerical interpolation for the missing value of {the 24th product} ``Metal Products, Machinery and Equipment Repair Services"  in order to get a complete curve of 42 products.
One can see from the figure the main differences among the the blue, black and red curves are that the {top} product in the red curve is {the 20th product}--``Communication Equipment, Computers and other Electronic Equipment", while the {top} product in the blue cure is {the 12th product}--``Chemical Products".
It is not difficult to understand why such a change occurred between 2012  and 2017,
because this period is the time mobile phones and other electronic devices began to become popular.
If we take the blue curve of 2012 as the benchmark, mark the top 6 products in a decreasing order with circled numbers, the top 6 of the blue curve and that of the black curve are basically the same except for the 5th one.

One can also see from Figure \ref{fig:exam6.1}
that the rankings of the 30th product ``Transportation",
{the 33th} ``Finance", the 34th ``Real Estate" and the 35th ``Leasing and Business" in
these three curves gradually increase.
We also find that, contrary to traditional popular belief,  finance and real estate are not pillar products.

\begin{figure}[h]
\begin{center}
\includegraphics[width=12.8cm,height=14.0cm]{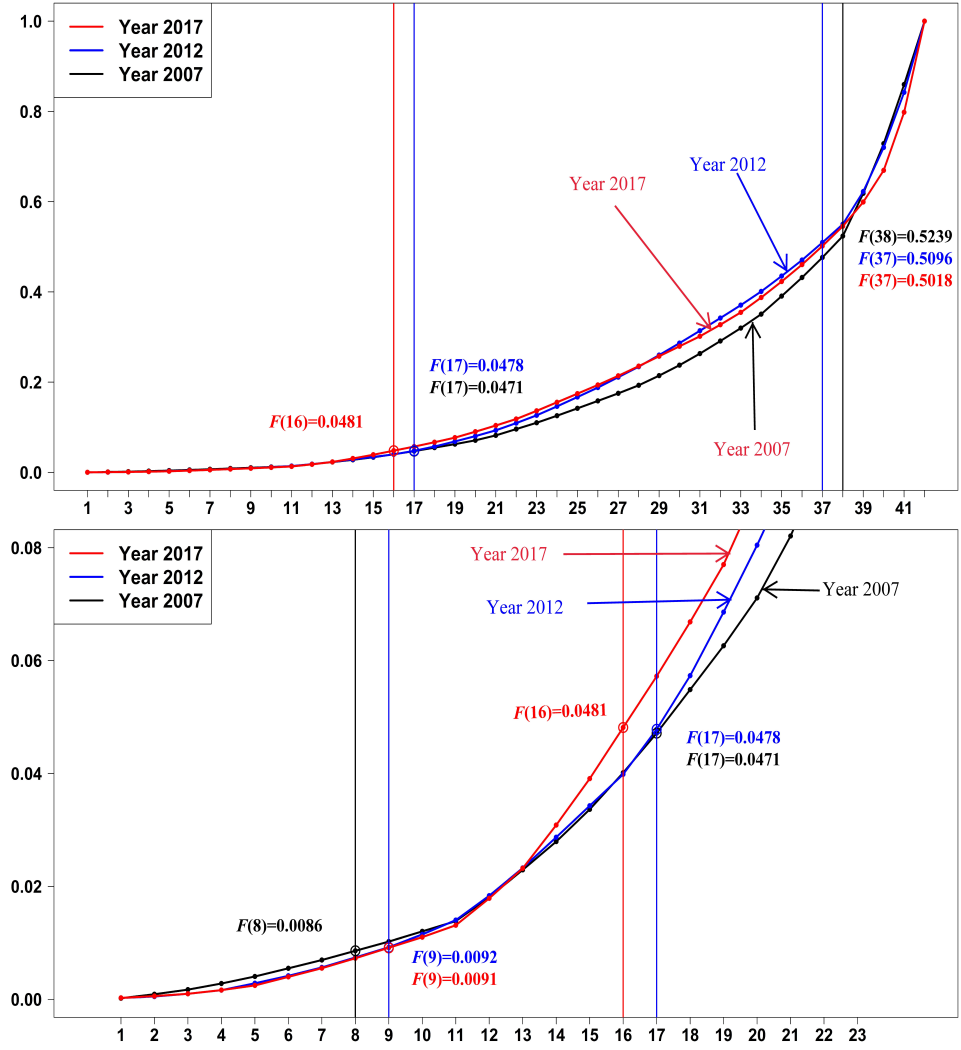}
\end{center}
\caption{Cumulative distribution of equilibrium solutions of 42 products in 2017 and 2012}
\label{fig:exam6.2}
\end{figure}

Figure \ref{fig:exam6.2} is the distribution function diagram generated by $\pi$.
Arranging the components of $\pi$ in ascending order $p_1<p_2<\cdots<p_{42}$,
we  get the discrete cumulative distribution function as follows:
$p_1$, $p_1+p_2, \cdots,\sum_{k=1}^{42}p_k$.

We plot all three  cumulative distribution graphs in Figure \ref{fig:exam6.2}.
There are two graphs in Figure \ref{fig:exam6.2},
the upper graph is the full graph, and the lower graph is a partial {one}.
The black, blue and red curves are the cumulative distribution {curves} of 2007, 2012 and 2017 respectively.
The horizontal axis in Figure \ref{fig:exam6.2} is the rank order of each product,
the vertical axis is the cumulative distribution of the first $n$ products.
To the left of the left vertical line,  the cumulative distribution is less than or equal to  5\%, and to the right of the right vertical line, the cumulative distribution is larger {than} or equal to 50\%.  Products to the left of the left vertical line are the weak products, Products to the right of the right vertical line are the pillar products, products between the two vertical lines are the intermediate products.
 One can see from Figure \ref{fig:exam6.2} that  there are 17 weak  products and 5 pillar products in 2007. One can also see that the cumulative distribution curves spanning 15 years are very consistent, indicating that using cumulative distribution to classify product grades is reliable.
Since the number of weak products is relatively large, we further divided them in the lower graph. One see that there 8 or 9 products with cumulative distribution {is} less than or equal to 1\%.

Figure \ref{fig:exam6.1} and \ref{fig:exam6.2} are taken from \cite{ycz2024}.

\section{{Quantum wave probability}}
The materials of this section are mainly {taken} from Chen's paper  \cite{c2024}.
Consider the complex matrix  $A$.
There are some extended versions of the Perron-Frobenius theorem, but the results are quite limited.
However, a point in Section 5, the transform from $A$ to $P$, still makes sense.

\begin{definition}
A complex matrix $A$ is called an SR1 matrix if $A{\bbb{1}}={\bbb{1}}$, i.e.,
the row sums are all equal to 1.
\end{definition}

Assume $\lambda\ne 0$  is a simple eigenvalue of the complex matrix $A$.
Similar to Chen's theorem on the key transform in Section 5, we have the following result.

\begin{theorem}[Generalized Chen's Theorem]\label{the:Gchen}
 Suppose that $(\lambda,u)$ and $(\lambda,v)$  are
{left- and right-eigenpairs} of the complex matrix
 $A$, respectively, $uA=\lambda u$, $Av=\lambda v$, and
 all components of $v$ are non-zero.
 For a column vector $w$ with all its components {being} non-zero, define
$$
R_w=D^{-1}_w\frac A\lambda D_w.
$$
Then the following conclusions hold:
\begin{itemize}\setlength{\itemsep}{-0.8ex}
\item[{\rm(1)}]
$R_w$ is an SR1 matrix if and only if $w=v$;
\item[{\rm(2)}] $R_v$ has a {left-eigenpair}
$(\lambda,u\odot v)$: $(u\odot v)R_v=\lambda(u\odot v)$.
\end{itemize}
\end{theorem}

For a Hermitian matrix $A$, {left- and right-eigenvectors} of $A$ corresponding to the same eigenvalue are conjugates of each other,  so  $u=\bar v$. Thus by Theorem \ref{the:Gchen},
 the {left-}eigenvector of $R_v$ is $u\odot v=\bar v\odot v$. Hence, we have the following corollary.

\begin{corollary}
If in the theorem above we further assume that $A$ is Hermitian,
then the {left-}eigenvector of $R_v$ is $\bar v\odot v$.
\end{corollary}


We know that
a discretized Schr\"odinger operator
corresponds to a Hermitian matrix, and its eigenvalues correspond to discrete energy levels. In quantum mechanics,
the square of the modulus of  the wave function
is the probability density of particle distribution, that is:
\begin{eqnarray*}
\pi=\frac{\bar v\odot v}{|v|^2}=\frac{\left(|v_1|^2,|v_2|^2,\cdots,|v_d|^2\right)}{|v|^2}.
\end{eqnarray*}

In other words, the vector $\bar v\odot v$
contains three characteristics of the Hermitian matrix $A$:
the {maximal} eigenvalue $\lambda$ and the corresponding {left- and right-eigenvectors} $u$ and $v$.
Since $\bar v\odot v$ represents the square of
the modulus of  $A$'s wave function
(equivalently, the eigenvector) $v$, this explains why
in Born's comment about matrix mechanics one should use ``the square of the modulus"  instead of ``the modulus".
{Once $|v| {<} \infty$, one} can use its normalized probability measure $\pi$ instead of $\bar v\odot v$ to
denote the probability density of particle distribution.
In summary, this is just an explanation of the same thing in two different languages, and {so} there is no objective randomness, just as Einstein said, ``God does not play dice".

Note that the real symmetric matrix is a special case of the complex Hermitian matrix, so the above conclusions {can also} be applied to {the} principal component analysis.

\section{{Economic forecasting and adjustment}}
In practice, for a model with consumption,  to determine a suitable economic growth rate, one needs to go through a {number} of adjustments and computations.
For a predetermined growth rate, we
first compute the consumption parameters, and then the available consumption.
The program goes {as} follows:
\begin{align*}
  \text{Growth rate} & \Rightarrow \text{Consumption parameters} \\
  & \Rightarrow \text{Available consumption} \overset{?}{\ge} \text{Planned consumption}
\end{align*}

If ``${\ge}$" holds, then the system can achieve the predetermined speed, and one can stop.
Otherwise, one can test again by reducing the growth rate. One can use  the optimization method to design these tests.

\begin{theorem}[\cite{c2022}]
 If the economic growth rate {satisfies}
 { $0<\delta<\min\{1,\rho(A)^{-1}-1\}$},
 {then} the available consumption in the
 $(n+1)$-th year is
\begin{align}\label{8.1}
\xi_n=\frac{1-(1+\dz)\rz (A)}{\dz}(x_{n+1}-x_n),
\end{align}
where $\{x_n\}$ is the iterative sequence {with given} {the} initial value {$x_0$} under $A_\alpha$ in Chen's model:
$${x_n=x_{n+1}A_\alpha,\qquad \alpha:=\frac{(1+\delta)^{-1}-\rho(A)}{1-\rho(A)}.}$$
Conversely, one can use the fact the consumption in the $(n+1)$-th year  does not exceed $\xi_n$ to determine {the} maximal growth rate $\delta$.
\end{theorem}

\nnd{ \bf Proof:} According to Theorem \ref{the:hua} (i),
the economic growth rate (denoted {by} $\delta$) is
$1/\rho(A_\alpha)-1$,
so
\begin{align}\label{8.2}
\delta=\frac 1{\rho(A_\alpha)}-1=\frac 1{(1-\alpha)\rho(A)+\alpha}-1.
\end{align}
Solving the equation \eqref{8.2}, we  get the consumption parameter
\begin{align}\label{8.3}
\alpha=\frac{(1+\delta)^{-1}-\rho(A)}{1-\rho(A)}=:\alpha(\delta)
\end{align}
and {the} consumption multiple
\begin{align}\label{8.4}
\gamma_\alpha=\frac\alpha{1-\alpha}=\frac{(1+\delta)^{-1}-\rho(A)}{1-(1+\delta)^{-1}}=:\gamma(\delta).
\end{align}
Therefore,
the available consumption {in} the $(n+1)$-th year is
$$\xi_n(\alpha)=\gamma_\alpha(x_{n+1}-x_n)=\frac{(1+\delta)^{-1}-\rho(A)}{1-(1+\delta)^{-1}}(x_{n+1}-x_n).$$
That is, \eqref{8.1} holds.
\qed

Combining \eqref{8.3} and \eqref{8.4} with $\rho(A)<1$, we get that $\alpha(\delta)$ and $\gamma(\delta)$ are both monotonically decreasing functions of $\delta$.
This implies that {the} consumption decreases when the economic growth rate increases, consumption increases when the economic growth rate decreases.

It can seen from the proof above that, if the economic growth rate $\delta$ is given, then
we can get $\alpha$ from \eqref{8.3} and then
$A_\alpha$.
{Regarding} the d-vector $x_n$ {as input}, one can use Chen's model {$x_n=x_{n+1}A_{\az}$} to find the output d-vector $x_{n+1}$ of the $(n+1)$-th year, and then get {the} available consumption of the $(n+1)$-th year.

Conversely, if the available consumption of the{ $(n+1)$-th year} and the input d-vector $x_n$ are given, the core of predicting the economic growth rate $\delta$ lies in to solve the consumption parameter $\alpha$. For this we have the following result.

\begin{corollary}
If the actual required consumption $($planned consumption$)$ is $\bar\xi_n$, then we only need to find $\alpha$ so that $\bar\xi_n\le\xi_n(\alpha)$.
Naturally, we define $\bar\alpha=\inf\{\alpha\in(0,1):\xi_n(\alpha)\ge\bar\xi_n\}$.
$\bar\alpha$ is the consumption parameter corresponding to $\bar\xi_n$.
Since $\xi_n(\alpha)$  increases with $\alpha$, we have, for any $\alpha\ge\bar\alpha$,
$$
\xi_n(\alpha)\ge\xi_n(\bar\alpha)\ge\bar\xi_n.
$$
Then the planned consumption $\bar\xi_n$ is alway less than or equal to the available consumption
$\xi_n(\alpha)$. $\delta(\bar\alpha)$ can be determined using {the} definition of $\bar\alpha$ and \eqref{8.2}.
\end{corollary}

Given $x_n$ and $A$, the available consumption $\xi_n(\alpha)=\gamma_\alpha(x_{n+1}-x_n)$  can be uniquely determined by $\alpha\in(0,1)$.
In fact, $\xi_n(\alpha)$ is {proportional} to
$x_{n+1}-x_n$.
More precisely,
the ratio of these two is $\alpha/(1-\alpha)$,
this is a carefully designed special quantity.
In general, one cannot expect that an arbitrary planned consumption $\bar\xi_n$ will be equal to the available consumption $\xi_n(\alpha)$ for a certain $\alpha$.
However, such {a} strong {restriction is} not really needed in practice,
because it is not always necessary to reduce economic growth to increase consumption.
One can adjust the consumption structure or imports to achieve this.

\section{Product regulation and optimization of economic structure}

By ``Product regulation" we mean to determine the products or industries that should be prioritized (or eliminated) in order to
adjust and optimize  the economic structure. This should be based on the current economic situation, needs rigorous computation and analysis.

This is a very difficult problem.
 Chen raised this as an open problem in 2005 (see \cite{{c2005}}). It was not until 2022 that he completely solved this problem.
The biggest challenge is to find the optimization criteria.
In traditional optimization theory, one tries to find the values of certain variables (under certain constraints) so that the chosen cost function attains the optimum. The problem is that, in our economic model, {we} do not know what the cost function is.

In our economic model {here, we have actually done two times for the optimization}: one is to achieve {the} fastest growth rate and the other is make sure  collapse never happens.
Hua's fundamental theorem tells us that the unique solution of both optimizations is the {maximal} {left-eigenvector} $u$ of $A$.
This  prompts us to take the d-vector ${\tilde u}=({ \tilde u}_k)$ as the target.
In other words, {first, we should find a}  target $\tilde u$ for product regulation.
Next, we construct a new structure matrix ${\tilde A}$ with ${\tilde u }$ being its {maximal} {left-}eigenvector.
This is a very complicated problem.
For example, if we consider a system with 100 products, we need to determine
the $10^4$ elements of the new {structure} matrix.
However, the given new {equilibrium} only provides $100$ data points. Where can we find the remaining $9900$ data points?  In order to fill in the missing information of the new matrix, we can require the new structure matrix ${ \tilde A}$ to be as close as possible to the original structure matrix $A$.


To complete the construction of the new structure matrix, we first make some necessary preparations in the next section, and then we return to the topic in Section 11.

\section{Markov chain, dual Markov chain and invariants of economic structure matrix }

Recall that Chen's transform ${\tilde A}$ of the matrix $A$ is defined by
\begin{align}\label{10.1}
\frac{{\tilde A}}{\rho\big({\tilde A}\big)}=D_w^{-1}\frac{A}{\rho(A)}D_w,\qqd w>0.
\end{align}
It follows from Theorem \ref{the:chen1989}(i)  that ${\tilde A}$ is a transition probability matrix $P$ if and only if $w$ is the {maximal right-eigenvector} $v$ of $A$.
We will often call $P$ a {\bf Markov chain}.
The { \bf three major characteristics} ({the} maximal eigenvalue and the corresponding {maximal} {left- and right-eigenvectors}) of $P$ are ($1, u\odot v, \bbb{1}$).

Let $A^*$ be the transpose of $A$, and  $u^*$  the corresponding {maximal right-eigenvector} of $A^*$.
It follows from  \eqref{10.1} that
\begin{align}\label{10.2}
D_w^{-1}\frac{A^*}{\rho(A)}D_w,\qqd w>0,
\end{align}
gives another transition probability matrix if and only if $w=u^*$.
The three characteristics of \eqref{10.2} are $(1, u^*\odot v^*, \bbb{1})$ (of course, $u^*\odot v^*=u\odot v$).
To remove $*$ in the above formula, we {need to make} the transposition and {then} get
$$Q_w:= D_w \frac{A}{\rho(A)}D_w^{-1}. $$
Now the column sums of $Q_w$ are all equal to 1 if and only if $w=u$.
We call $Q_u$ the {\bf Dual Transition Probability Matrix} or {\bf Dual Markov chain} (abbrev.{\bf Dual Chain}).
Its three major characteristics are $(1, \bbb{1}^*, u\odot v)$.

Before studying the invariants of the economic structure matrix,
we first review the simplest invariant:
$$
\pi= \frac{\text{circumference of the circle}}{2\,r}=\frac{\text{area of the circle}}{r^2},
$$
which omits the basic parameter describing the size of a circle--radius $r$.
Recall our consumption model:
$$A_{\az}=(1-\az)A+\az I,\qqd \az\in [0,1).$$
Omitting the spectral radius $\rho(A_{\az})$ that describes the development speed,
this family of matrices has two invariants: they have the same {maximal} {left- and right-eigenvectors} $u$ and $v$.
Two of the three major characteristics become invariants (independent of $\az$).
We have seen that these two invariants
(vectors rather than scalars)
play an important role in economic theory.

Now for a given $P$,
 we consider the inverse transform of the transform \eqref{10.1}.
For any positive $w$, we can define {Chen's transform of $P$}
$$A_w:=D_w P D_w^{-1}.$$
Since $A_w$ is similar to $P$, we have $\rho(A_w)=1$.
It is easy to show that $w$ is the {maximal right-eigenvector} of $A_w$.
Therefore, by \eqref{10.1}, the transition probability matrix derived from $A_w$ is $P$.
This shows that the  family of square matrices
$$\mathscr{A}_P=\{A_w=D_w P D_w^{-1}: w>0\}$$
have the same {\bf invariant square matrix} $P$, which is a square matrix instead of a vector.

We can deal with the dual case similarly. For a fixed dual Markov chain $Q$,
we define the family of square matrices
$$\mathscr{A}_Q:=\{A_w=D_w^{-1} Q D_w: w>0\}.$$
For any $A_w\in {\scr A}_Q$, it is obvious that $\rho(A_w)=\rho(Q)=1$.
It is easy to show that $w$ is the {maximal} {left-}eigenvector of $A_w$, so the dual Markov chain derived from $A_w$ coincides with $Q$.
This says that the family of square matrices ${\scr A}_Q$ has $Q$ as an {\bf invariant} ({\bf invariant square matrices}).

The above invariant vectors or invariant matrices are collectively called {\bf Chen's  invariants}.

Note that since $w>0$,
we can replace $w$ by $w^{-1}$, if necessary.
In the definition of the set ${\scr A}_Q$,
$D_w^{-1} Q D_w$ becomes  $D_w^{\mp} Q D_w^{\pm}$, so the set ${\scr A}_Q$ is large enough.
This is useful in practice, because we often need to adjust $w={\tilde u}$.
Of course, in the definition of ${\scr A}_Q$,
we can replace $A_w$ by $A_w/\rho(A_w)$, to add a degree of freedom to ${\scr A}_Q$.

\section{Product regulation and optimization of economic structure~(continued)}

Now, we return  to the topic of Section 9:
construct a new structure matrix~$\tilde A$ with~$\tilde u$ as the equilibrium solution,
making it as close as possible to the original matrix $A$.
We are given $A$ and its {maximal} {left-}eigenvector $u$, and the target equilibrium solution $\tilde u$.
From the previous section, we have a completely determined dual chain
$$Q_u:= D_u \frac{A}{\rho(A)}D_u^{-1}. $$
It is in a one-to-one correspondence with $A/\rho(A)$:
$$\frac{A}{\rho(A)}=D_u^{-1}Q_u D_u.$$
Replacing $A$ with the desired ${\tilde A}$, we formally get two equations parallel to the equations above
$$Q_{\tilde u}:= D_{\tilde u} \frac{{\tilde A}}{\rho({\tilde A})}D_{\tilde u}^{-1},\qqd
\frac{{\tilde A}}{\rho({\tilde A})}=D_{\tilde u}^{-1} Q_{\tilde u}D_{\tilde u}.
$$
Since $\tilde u$ is given, ${\tilde A}$ and the corresponding $Q_{\tilde u}$ are the only unkowns in the last two equations.
To make ${\tilde A}$ as close to $A$ as possible, one should  require that their invariants to be as close as possible.
The simplest way to do this is let $Q_{\tilde u}=Q_u$.
Starting from~$Q_{\tilde u}$ and~$Q_u$, we can define, as we did in the previous section, two families of square matrices
with $Q_{\tilde u}$ and $Q_u$ as invariants:
$$\aligned
\scr{A}_{Q_u}=\bigg\{A: \frac{A}{\rho(A)}=D_w Q_u D_w^{-1},\; w>0\bigg\},\\
\scr{A}_{Q_{\tilde u}}=\bigg\{{\tilde A}: \frac{{\tilde A}}{\rho({\tilde A})}=D_w Q_{\tilde u} D_w^{-1},\; w>0\bigg\}.
\endaligned$$
Clearly, $\scr{A}_{Q_{\tilde u}}=\scr{A}_{Q_u}$ once $Q_{\tilde u}=Q_u$.
Thus we can find the only unknown quantity ${{\tilde A}}\big/{\rho({\tilde A})}$ from the above system of equations:
$$\aligned
\frac{{\tilde A}}{\rho({\tilde A})}
&=D_{\tilde u}^{-1} Q_{\tilde u}D_{\tilde u}\qd\text{(first take~$w=
\text{the known}\,{\tilde u}$ in~$\scr{A}_{Q_{\tilde u}}$)}\\
&=D_{\tilde u}^{-1} Q_{u}D_{\tilde u}\qd\text{(let~$Q_{\tilde u}=\text{the known}\,Q_u$)}\\
&=D_{\tilde u}^{-1} D_u \frac{A}{\rho(A)}D_u^{-1}D_{\tilde u}\qd\text{(plug in the explicit expression of $Q_{u}$)}\\
&=D_{{\tilde u}^{-1}\odot u}\frac{A}{\rho(A)}D_{u^{-1}\odot{\tilde u}}\qd\text{(simplify the right hand side)}\\
&=D_{{\tilde u}\odot u^{-1}}^{-1} \frac{A}{\rho(A)}D_{{\tilde u}\odot u^{-1}}.
\endaligned$$
Because~$\{A_{\az}, \az\in [0, 1)\}$ have the invariants~$u$ and~$v$, this result can be immediately extended to the case with consumption,
which leads to the following theorem.

\medskip

\begin{theorem}$($Chen's Economic Structure Optimization Theorem$)$
The optimal matrix $\tilde A_\az$, with {the} target $\tilde u$, of the structure matrices $A_\az$ with {the} {maximal left-eigenpair} $(\rho(A_\az), u)$ can be determined by \eqref{10.1} and $w={\tilde u}\odot u^{-1}$:
\begin{align}\label{11.1}
\frac{{\tilde A_{\az}}}{\rho({\tilde A_\az})}=D_{{\tilde u}\odot u^{-1}}^{-1} \frac{A_{\az}}{\rho(A_{\az})}D_{{\tilde u}\odot u^{-1}}, \qqd \az\in [0, 1).
\end{align}
Furthermore, the {maximal} {left- and right-eigenvectors} of $\tilde A_\az$ are $\tilde u$ and
${\tilde v}=v\odot u\odot {\tilde u}^{-1}$.
\end{theorem}
\nnd{ \bf Proof:}
We only need to show the last assertion.
Since~$\{A_\az: \az\in[0,1)\}$ have the same {maximal} {left- and right-eigenvectors}~$u$ and~$v$, we will omit $\az$ in this proof.
\medskip

\nnd (a)
First, we show that ${\tilde u}$ is the {maximal} {left-}eigenvector of ${\tilde A}$, that is,
$${\tilde u}\frac{{\tilde A}}{\rho({\tilde A})}={\tilde u}.$$
Note that for any given vectors $x$ and $y$, we have $xD_y=x\odot y=D_x y$. So
$$ {\tilde u}D_w^{-1}={{\tilde u}\odot w^{-1}}={{\tilde u}\odot {\tilde u}^{-1}}\odot u=u,$$
From~ \eqref{10.1}, we can conclude that
$${\tilde u}\frac{{\tilde A}}{\rho({\tilde A})}
= {\tilde u} D_w^{-1} \frac{A}{\rho(A)} D_w= u \frac{A}{\rho(A)}\,D_{w} = u D_w={\tilde u}.$$
This proves that ${\tilde u}$ is the {maximal} {left-eigenvector} of ${\tilde A}$.

\nnd (b)
Now we prove that $\tilde v$ is {the} {maximal right-eigenvector} of ${\tilde A}$.
 From the definition of $w$ and $\tilde v$, we know that
\begin{align}\label{11.2}
D_w{\tilde v}=w\odot {\tilde v}={\tilde u}\odot u^{-1}
\odot v\odot u\odot {{\tilde u^{-1}}}=v,
\end{align}
Hence
$$\frac{{\tilde A}}{\rho({\tilde A})}{\tilde v}
=D_w^{-1}\frac{A}{\rho(A)}D_w {\tilde v}=D_w^{-1}\frac{A}{\rho(A)}v=D_w^{-1} v={\tilde v}.\qed$$

We have proved that the optimal structure matrix defined by \eqref{11.1} has the predetermined target d-vector as the {equilibrium}.
With ${\tilde A}$ and $\tilde u$, we get a new economic system.
To understand the new system, classify products and test system stability,
it is natural to use the equilibrium state $u\odot v$ of the dual chain $Q_{\tilde u}$
(which was taken to be $Q_u$).
$u\odot v$ is the
{maximal right-eigenvector}  of the dual chain, and it is also the equilibrium solution of the invariant $P$ derived from  the original structure matrix.
Thus, ${\tilde A}$ retained the invariant $P$ from $A$.
In fact, to  classify products and test system stability for the optimal structure matrix ${\tilde A}$, we can simply return to $P$
derived from  the original structure matrix $A$, there is no need to use {the} dual chain $Q_{\tilde u}$.
The proof is as follows.
From the known $\tilde v$ and $w={\tilde u}\odot u^{-1}$,
we have proved in \eqref{11.2} that $w\odot {\tilde v}=v$, thus
$${\tilde P_\az}=D_{\tilde v}^{-1} \frac{\tilde A_\az}{\rho({\tilde A_\az})} D_{\tilde v}
=D_{\tilde v}^{-1} D_w^{-1}\frac{A_\az}{\rho(A_\az)} D_wD_{\tilde v}
=D_{v}^{-1}\frac{A_\az}{\rho(A_\az)} D_{v}
=P_\az.$$
In other words, we prove the following
{main} property (i) of the optimal matrix.
\medskip

\begin{theorem}
{
Given a target d-vector $\tilde u$ and $\az\in [0, 1)$, let $\tilde A$ be the optimal matrix determined by the above theorem.  Then
\vspace{-0.25cm}
\begin{itemize}\setlength{\itemsep}{-0.8ex}
\item[$\mathrm{(i)}$]
\vspace{0.2cm}
{the} invariant ${\tilde P}_{\az}$ of the optimal matrix ${\tilde A}_{\az}\big/\rho\big({\tilde A}_{\az}\big)$ coincides with the invariant $P_{\az}$ of the original matrix $A_{\az}/\rho(A_{\az})$.

\vspace{0.2cm}
\item[$\mathrm{(ii)}$]
{the} stability index (collapse time, collapse location) are the shared invariants of $A_{\az}, {\tilde A}_{\az}$ and $P_{\az}$.
\end{itemize}
}
\end{theorem}
\nnd{ \bf Proof:}
We only need to prove {the} assertion {(ii)}.
The stability of $A_{\az}, {\tilde A}_{\az}$ and $P_{\az}$ are all realized through $P_{\az}$. It follows from (i) that $A_{\az}, {\tilde A}_{\az}$ have the same invariant $P_{\az}$. Thus, to prove (ii), we only need to show that
$A_{\az}$ and~$P_{\az}$ have the  common stability index.
This can be derived the result below. More precisely, since we are concerned with when~(i.e.~$n$)~
and where~(i.e. which product) a negative sign first appears in $\mu_n$ and $x_n$, we can omit the positive factors $\rho(A_\alpha)^n$ and $v$ on both sides of the identity \eqref{11.3} below, then we get the desired assertion.
\qed
\medskip

The following result is a natural extension of Theorem \ref{the:mux}.
\begin{theorem}
\nnd{$($Conversion Theorem/Equivalence Principle$)$}~$($\cite{5rw24}: Theorem~$7.3$$)$\;{
The iterative sequence~$\{\mu_n\}_{n\ge 0}$ of $P_{\az}$, the iterative sequence~$\{x_n\}_{n\ge 0}$ of ~$A_{\az}$ and~$v$ satisfy the identity:
\begin{align}
&\mu_n=\rho(A_{\az})^n x_n\odot v,\qqd n\ge 0, \label{11.3}\\
&x_n= \rho(A_{\az})^{-n} \mu_n\odot v^{-1},\qqd n\ge 0. \label{11.4}
\end{align}
Therefore, the two algorithms are equivalent.
}
\end{theorem}

If we take $x_0=u$ in \eqref{11.3}, then $\mu_0=u\odot v$. Multiplying both sides of \eqref{11.3} by the vector $\bbb{1}$ on the right {side}
yields $\rho(A)^n x_n\odot v\bbb{1}=\mu_n\bbb{1}$.
In particular, when $n=0$, we get the normalization condition $uv=\mu_0\bbb{1}=1=\pi\bbb{1}$.
Note that $\mu_n$ and $x_n$ differ by an exponential {maximal} order $\rho(A)^n$ and a constant vector factor $v$.

The above theorem also applies to the general transform $A\to {\tilde A}$ defined by \eqref{10.1}.

\section{Programmable and intelligentizable efficient algorithms }

The main difference between  Hua's economic optimization theory and other existing economic theories is that Hua's theory is computable and programmable.
One of the important cornerstones of Chen's new economic optimization theory is the computation of the
{maximal} eigenvalue of a nonnegative matrix and the corresponding {maximal} {left- and right-eigenvectors}.
For system stability analysis, we not only need these three quantities, we also need to know how {to} compute them
accurately and efficiently.

Efficient algorithms play a significant role in Chen's new theory.
In recent years, by using stochastics,  Chen made many important progresses in efficient  algorithms for matrix eigenpairs.
It is generally believed that power method converges too slowly and is of little use and that, the  inverse power method, due to its use of the Rayleigh entropy estimate, may not be reliable. Chen introduced a new security estimate and completely avoided using the Rayleigh entropy estimate. By {combining} ideas from machine learning with these two methods and their variants, he achieved an universal high-speed algorithm.

Using our {present computer} and existing algorithms, it is not difficult to compute the {maximal} eigenvalues and the corresponding {maximal} {left- and right-eigenvectors} of low dimensional or symmetric matrices. However, accomplish the same task for high-dimensional non-symmetric matrices is still very daunting.
To address this problem, Chen proposed effective algorithms: {\bf Matrix quasi-symmetrization technique} and {\bf Eigenvector smoothing technique}. The former is used to reduce the amplitude of the matrix ($\max A-\min A=\max_{ij}a_{ij}-\min_{ij}a_{ij}$), and the latter is to make the right eigenvector $v$ as flat as possible, i.e., {$\max v/\min v=\max_iv^{(i)}/\min_iv^{(i)}$} is nearly constant.

\vspace{3mm}
\nnd{\bf{Matrix quasi-symmetrization technique}\,(cf.\,[\ref{cl19}];\,\S3):}
Given a nonnegative irreducible matrix $A=(a_{ij})$,
we define the following $Q$ matrix (a matrix with nonnegative off diagonal elements and {having zero-sum of each row}):
$$Q=A-D_{A\bbb{1}}.\vspace{-0.2truecm}$$
Using irreducibility, we know that the equation
$$\mu\, Q=0$$
has a unique positive solution $\mu=(\mu^{(1)},\mu^{(2)},\cdots,\mu^{(d)})$ satisfying the initial condition $\mu^{(1)}=1$.
With this positive vector $\mu$, we can define the {\bf quasi-symmetrized} matrix ${\hat A}$ of $A$:
$${\hat A}=D_{\mu^{1/2}} A\,D_{\mu^{-1/2}}.$$
The important reason for introducing this concept is that  $A$ is {{symmetrizable}} with respect to  $\mu$, that is,
$$D_{\mu} A =A^* D_{\mu}\qd[=(D_{\mu} A)^*].$$
It is easy to check that $A$ is symmetrizable with respect to  $\mu$ iff $\hat A$ is symmereic. Hence this is a very important generalization of symmetry.
\cite{cmf18} illustrated the irreplaceable power of the quasi-symmetrization technique for computing the eigenpairs of
non-symmetric but symmetrizable matrices through a simple tridiagonal matrix. We can not expect a general $A$
to be symmetrizable, but $\hat A$ often reduces the amplitude of $A$.

\vspace{3mm}
\nnd{\bf Eigenvector Smoothing Technique\,(cf.\,[\ref{cl19}];\,\S4)}:
Both the power method and the inverse power method are mainly for computing eigenvectors of matrices.
If the amplitude of the eigenvector is too large,
it is simply impossible to achieve this using these two methods.
Therefore, further transforming $\hat A$ into a matrix with as {a flat} eigenvector as possible (i.e., $\max v/\min v$ is nearly constant) is naturally beneficial for computation. At first glance, this problem seems difficult to approach, but with Chen's Theorem \ref{the:chen1989}, the desired conclusion comes naturally.

\medskip
\begin{theorem} { Let $w$ be any positive vector, and define
$$\bar A:=A_w=D_w^{-1} A\,D_{w}=(w^{-1}\otimes w) \odot A,$$
where { $w^{-1}\otimes  w=\left({w^{(j)}}/{w^{(i)}} \right)_{d \times d}$}
is the tensor product of vector $w ^ {-1} $and vector~$w $. Then $A_{\bbb{1}}=A$. Denote the
{maximal right-eigenpair} of
$A_w$ as $(\rho(A_w), g_w)$.
Then we have
$$\rho(A_w)=\rho(A),\qqd g=g_{\bbb{1}}=D_{w} g_w.$$
In particular, if $\max_i |w^{(i)}-v^{(i)}|$ is sufficiently small, then $g_w$ is a vector close to a constant.}
\end{theorem}
\nnd{\bf {Proof:}}
Since $A_w$ is similar to $A$, it shares the same eigenvalues with $A$,
and the eigenvector $g_w$ satisfies
$$\rho(A)g_w=D_w^{-1}AD_wg_w.$$
Thus, we have $\rho(A)D_wg_w=A(D_wg_w)$, and hence the first assertion holds.
Now, we prove the last assertion. Since
$$D_wg_w=g=D_vg_v=D_v\bbb{1},$$
we have
$$g_w=D_w^{-1}D_v\bbb{1}=D_{w^{-1}\odot v}\bbb{1}=w^{-1}\odot v.$$
This leads to the desired assertion, as $w^{-1}\odot v$ is a vector close to a constant.
\qed

Now we can state our eigenvector smoothing technique. Suppose $w$ is an approximate solution of $v$, then the
{maximal right-eigenvector} of $\bar A$ is nearly a constant vector.

\vspace{3mm}
Chen's transform (the key transform $A\to P$) plays a central role in {the} theory. Altogether, 7 {times are} used
in Chen's theory:

\begin{itemize}\setlength{\itemsep}{-0.4ex}
\item[(\romannumeral1)]~ {The Proof of Hua's Fundamental Theorem in {sections} {3 and 5};}
\item[(\romannumeral2)]~{In stability testing, he used $P$ instead of $A$ to study the stability of the system;}
\item[(\romannumeral3)]~{In product grading and classification, he used the {equilibrium} solution of $P$, i.e., the component product of the {maximal} {left- and right-eigenvectors} of $A$, which has a clear economic significance;}
\item[(\romannumeral4)]~{He uses the dual chain twice in economic structure optimization, i.e., he uses the key transform twice;}
\item[(\romannumeral5)]~{The stability test of
 the new optimal economic structure ${\tilde A}$
 returns to $P$ (${\tilde P}=P$);}
\item[(\romannumeral6)]~{The stability test of the new economic structure $\kappa {\tilde A}$ after {the} secondary optimization still uses $P$, $A$, ${\tilde A}$, and $\kappa {\tilde A}$ have exactly the same stability as $P$.}
\end{itemize}

From above, it can be seen that the transition probability matrix $P$ is an invariant undoubtedly.

\section*{Acknowledgments}
The second-named author first met Prof. Chen in the summer of 1988 at the ``Conference on Multi-parameter Stochastic Processes''  held at Qingdao University.  Since then,  Chen has given us constant encouragements.  He
has been a great inspiration to all of us. Since he moved to Jiangsu Normal University in 2019, he patiently guided us in our researches, especially in economic optimization theory and its applications. We all benefited greatly from his guidance. On the occasion of his 80th birthday,  we use this survey as token to show our appreciation to Chen.
We wish  Chen continued success in his research endeavors.
We would also like to thank Prof. Renming Song from University of Illinois Urbana-Champaign, Prof. Xiaoying Liu and Dr. Lianyong Qian from Jiangsu Normal University, Dr. Yueshuang Li from Capital University of Economics and Business for their help in revising.

\medskip

\smallskip
{\small
\nnd For popular videos (In Chinese. The English subtitles are adding) about the new theory, please refer to the following website.  http://rims.jsnu.edu.cn/f7/48/c16902a390984/page.htm\\
\nnd For video of lectures, see Chen's homepage: http://math0.bnu.edu.cn/~chenmf/ \\
\nnd{\bf Videos of Lectures}\qd [19]--[21].}
\end{document}